\documentclass[twocolumn,prl,amsmath,amssymb,amsfonts,superscriptaddress,floatfix,showpacs]{revtex4}

\usepackage{graphicx}
\usepackage{dcolumn}
\usepackage{bm}
\usepackage{rotating} 

\begin{document}

\title{
Accurate electronic band gap of pure and functionalized graphane from GW calculations
}
\date{\today}
\author{S.~Leb\`egue}
\affiliation{
 Laboratoire de Cristallographie, R\'esonance Magn\'etique et Mod\'elisations (CRM2, UMR CNRS 7036)
 Institut Jean Barriol, Nancy Universit\'e
 BP 239, Boulevard des Aiguillettes
 54506 Vandoeuvre-l\`es-Nancy,France
 }
\author{M.~Klintenberg}
\affiliation{
Department of Physics and Materials Science, Uppsala University,
Box 530, SE-75121, Uppsala, Sweden
 }
\author{O.~Eriksson}
\affiliation{
Department of Physics and Materials Science, Uppsala University,
Box 530, SE-75121, Uppsala, Sweden
 }
\author{M.~I.~Katsnelson}
\affiliation{
 Institute for Molecules and Materials, Radboud
University Nijmegen, Heyendaalseweg 135, NL-6525 AJ, Nijmegen, The
Netherlands }

\begin{abstract}
	Using the GW approximation, we study the electronic structure
	 of the recently synthesized hydrogenated graphene, named graphane.
	 For both conformations,
	  the minimum band gap is found to be direct at the $\Gamma$ point,
	   and it has a value of $5.4$ eV in the stable chair conformation, where H atoms attach C atoms alternatively on opposite sides of the two
	   dimensional carbon network. In the meta-stable boat conformation the energy gap is
	   $4.9$ eV.
	   Then, using a supercell approach, the electronic structure of graphane was modified by introducing
	    either an hydroxyl group or an H vacancy. In this last case, an impurity
	     state appears at about $2$ eV above the valence band maximum.
\end{abstract}


\maketitle
After the discovery of graphene\cite{first} and of its
extraordinary electronic properties\cite{r1,r2,r3} the chemical
functionalization of graphene has become the focus of special interest
in contemporary materials science. Being first a truly
two-dimensional crystal and demonstrating high electron
mobility\cite{first}, graphene is ideal for modern electronics
which is essentially two-dimensional\cite{r1}. At the same time, the
peculiar gapless ``ultrarelativistic'' energy spectrum of
graphene\cite{r2,r3} makes the creation of a ``carbon
transistor'' based on $p-n$ junctions highly nontrivial,
due to the anomalous transparency (``Klein
tunneling''\cite{klein}). Nanoscale graphene single-electron
transistor has been demonstrated already\cite{ponomarenko} but it
has a relatively restricted domain of potential applications.
Therefore, to transform graphene into a semiconductor with a
conventional electron spectrum keeping its two-dimensionality is a
real challenge in the field, and chemical functionalization is
considered as one of the most promising ways to solve the problem.

A new derivative of graphene was recently synthesized\cite{elias},
where hydrogenation turned graphene into what now is called graphane.
Previously this compound was studied theoretically, and it is actually an excellent example of a 
new and interesting material which was predicted from first principles theory\cite{graphane1,graphane2} before it was synthesised experimentally.
In the calculations of the electronic structure it was found that graphane is a semiconductor with a rather wide energy
gap\cite{graphane1,graphane2}. In the experiment\cite{elias} the
adsorption of hydrogen on graphene was indeed observed to result in that
a gap opened up in the electron states. Hence the adsorption of
hydrogen turned the highly conductive graphene into insulating
graphane, in accordance with the theoretical predictions.\cite{graphane1,graphane2}
However, the exact value of the band gap is still
experimentally unknown. Previous theoretical calculations giving
values from 3.5 eV\cite{graphane1} to 3.8 eV\cite{danil} use a
standard density functional with generalized gradient
approximation (GGA). It is well known that, in general, this
approach is not reliable to calculate energy gaps in
semiconductors often giving inaccurate results (the so called
``gap problem''\cite{godby}). At the same
time, knowledge about the size of this gap is crucial in order to
assess the possibility of using graphane in electronics
applications and for use in devices. 

The reason for a band gap opening up when hydrogen is adsorbed on
graphene is that sp$^2$ bonded C atoms become sp$^3$ bonded atoms,
where three of the four covalent sp$^3$ bonds are saturated by C
atoms and the fourth covalent bond is saturated by an H atom. This
is for instance illustrated in the theoretical part of
Ref.\onlinecite{elias} and is also known from the adsorption of
single hydrogen atoms on graphene\cite{wessely}. In accordance to
expectations this change in chemical binding is also reflected in
the electronic structure, where sp$^3$ bonded C atoms do not
display a so called $\pi^{*}$ peak in the x-ray absorption
spectrum\cite{wessely}. Hence the basic reason for the drastic
change in conductivity is in accordance to expectations for sp$^3$
bonded carbon (i.e. diamond), with an insulating behaviour. There
is of course no reason to expect that the size of the band gap of
graphane should be similar to that of diamond, even though there
are similarities in the nature of the chemical binding. Hence
measurements or accurate theoretical calculations are needed, and
in this article we present a calculation based on the GW
approximation. 


Due to the well-known deficiencies of DFT\cite{Hohenberg} to treat excited states, (for example the band gaps in the
 LDA or GGA approximations are much smaller than the experimental values), we have used the GW approximation (GWA)
 of Hedin\cite{Hedin1,Hedin2} to study the electronic structure of graphane.
In this formalism, the Kohn-Sham equations\cite{Hohenberg} are replaced by the quasiparticle equation:\\
$(T+V_{ext}+V_{h})\psi_{{\bf k}n}({\bf r}) + \int d^3r^{\prime}
 \Sigma({\bf r},{\bf r}^{\prime},E_n({\bf k}))\psi_{{\bf k}n}({\bf r}^{\prime}) $
\begin{eqnarray}
= E_n({\bf k})\psi_{{\bf k}n}({\bf r}) ~~~~~~~~ \nonumber
\label{eq:qp_psi}
\end{eqnarray}
where $T$ is  the free-electron kinetic energy operator, $V_{ext}$ the external potential
 due to the ion cores, $V_{h}$ the Hartree potential,
 $\Sigma$ the self-energy operator, and $E_n({\bf k})$ and $\psi_{{\bf k}n}({\bf r})$
 are respectively the quasiparticle energy and wave function.

 An adequate approximation for the self-energy operator is to write it as
  the product of the Green's function and the screened Coulomb interaction $W$,
   which yields to the so-called GW approximation.
 \begin{equation} \label{self_energy}
 \Sigma({\bf r},{\bf r}^{\prime},\omega)=\frac{i}{2\pi}\int d\omega'
 G({\bf r},{\bf r}^{\prime},\omega+\omega^{\prime})e^{i\delta\omega^{\prime}}
 W({\bf r},{\bf r}^{\prime},\omega^{\prime}) \nonumber
 \end{equation}
 Then, since the difference $\hat{\Sigma}-\hat{V}_{xc}$ between the self-energy and the Kohn-Sham
exchange and correlation potential is small, a perturbation theory is used to write the
QP Hamiltonian:
\begin{equation}
\hat{H}^{qp} = \hat{H}_{KS} + (\hat{\Sigma}-\hat{V}_{xc}) \nonumber
\label{eq:hqp}
\end{equation}
Finally, the  QP energies are obtained by expanding the real part of self-energy  to first order  around
$\epsilon^{DFT}_n({\bf k})$  and assuming that the the QP wave function $\psi_{{\bf k}n}$ and Kohn-Sham
 wave function $\Psi_{{\bf k}n}$ are identical: \\
$\textrm{Re}E_n({\bf k})) = \epsilon^{DFT}_n({\bf k})+ Z_{n{\bf k}} \times$
\begin{eqnarray}\label{quasiparticle_energy_final}
[\langle\Psi^{DFT}_{{\bf k}n}|
\textrm{Re}\Sigma({\bf r},{\bf r}^{\prime},\epsilon_n({\bf k}))|\Psi^{DFT}_{{\bf k}n}\rangle
- \langle\Psi^{DFT}_{{\bf k}n}|V_{xc}^{DFT}(r)|\Psi^{DFT}_{{\bf k}n}\rangle] \nonumber
\end{eqnarray}
where the QP renormalization factor $Z_{n{\bf k}}$ is given by
\begin{equation}\label{Renormalization}
Z^{DFT}_{n{\bf k}}=[1-\langle\Psi^{DFT}_{{\bf k}n}|
\frac{\partial}{\partial\omega} \textrm{Re}\Sigma({\bf r},{\bf r}^{\prime},
\epsilon_n({\bf k}))
|\Psi^{DFT}_{{\bf k}n}\rangle]^{-1}. \nonumber
\end{equation}
Therefore, it appears that the DFT eigenvalues $\epsilon_n({\bf k})$ are
 corrected by the GW approximation, giving a practical scheme to compute
 reliable excited state properties\cite{review_gw1,review_gw2, review_gw3}.

 Here we have used the code VASP\cite{vasp} (Vienna Ab-initio simulation package),
 implementing the projector augmented waves (PAW) method\cite{Bloechl} to
   compute the ground state and excited state properties of graphane.
   First, we obtained a reliable geometry of the structure by optimizing it
   for the two possible conformations (see below), using the generalized gradient approximation\cite{gga}. 
   Contrary to Sofo et al.\cite{graphane1},
     we have chosen to simulate completely isolated sheets of graphane, although
      the resulting bandstructure is very close to that other\cite{graphane1}. For all the calculations, we
       have used the default cut-off for the wavefunction. During the optimization
       of the structures, a k-point grid\cite{Monkhorst} of $10 \times 10 \times 1$ was used.
       Then, using the relaxed geometries, a final run was performed with a $16 \times 16 \times 1$ grid.
      Finally, these ground state calculations  were used to compute the quasiparticle bandstructure with VASP,
      following the method described in Ref. \onlinecite{gwkresse}.
	Two hundred bands were used for the summation over bands in the calculation of the polarisability
	 and the self-energy, and a cut-off of $150$ eV was used for the size of the polarisability matrices.
	 For details about the implementation of the GW approximation within the PAW formalism, see also Refs.
	 \onlinecite{brice,seb}.
	 Notice that the GW approximation was applied successfully to study the properties of graphene nanoribbons
	 \cite{louie1,louie2}, of graphene\cite{gwgraphene}, and of carbon nanotubes\cite{gwtubes}.

\begin{figure}[ht]
 \begin{center}
    \includegraphics[width=0.4\textwidth, angle=0.]{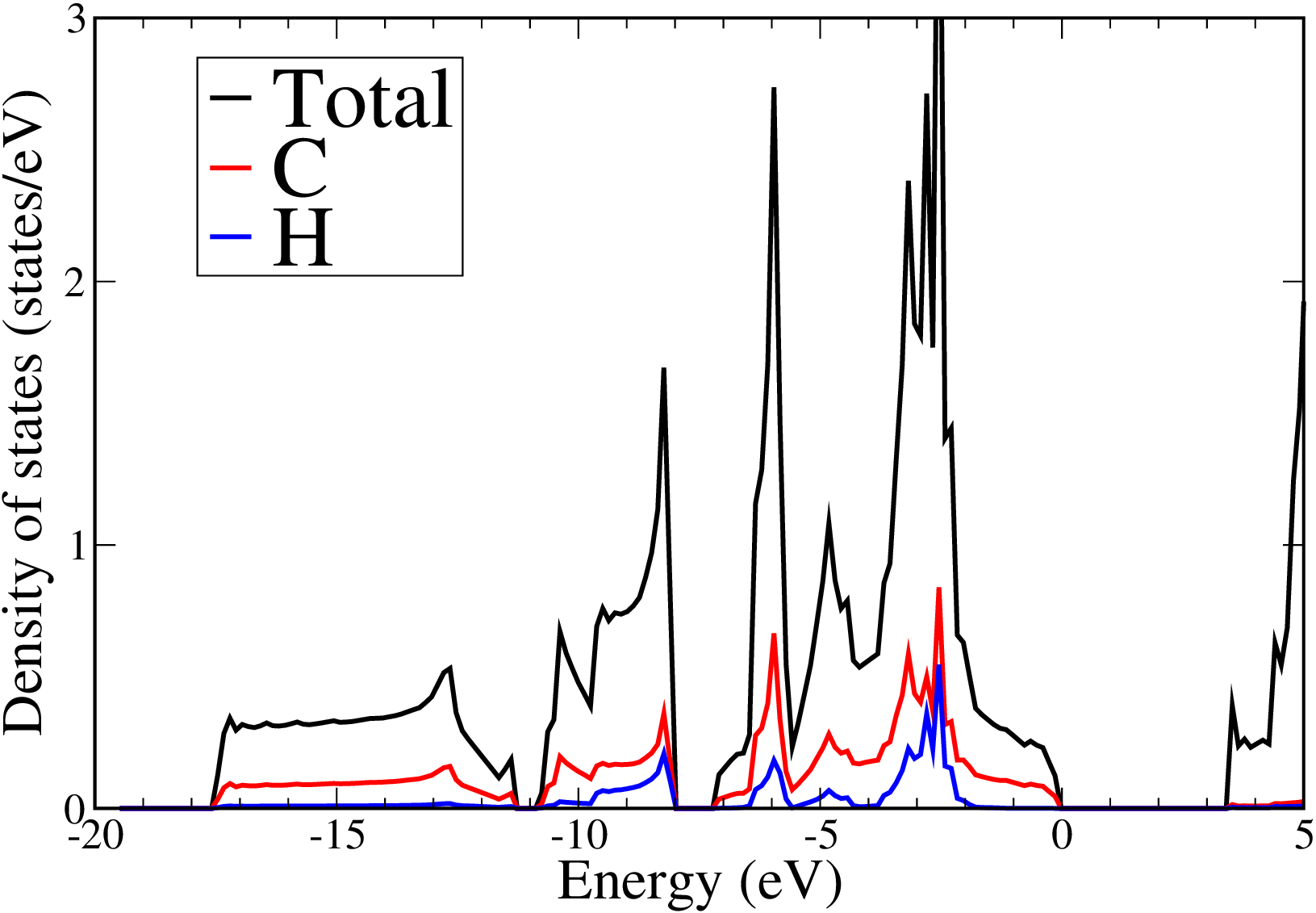}
    \caption{(Color online)
    The total (in black) and partial (in red for carbon and in blue for hydrogen) density of states of
     graphane in the chair conformation.
     The top of the valence bands is chosen as the zero energy.
 }
    \label{fig:partial}
  \end{center}
\end{figure}

Graphane is made of a hexagonal network of carbon
 atoms, with the orbitals being of the sp$^3$ kind and
  bonded to hydrogen atoms, with a carbon/hydrogen ratio of one.
  It exists in two conformers\cite{graphane1}, the chair-like conformer and the boat 
   like conformer. In the chair conformer, all C-C bonds are equivalent and 
 the hydrogen atoms are bonded alternatively on each side of the carbon plane (see Fig.1 of Ref.\onlinecite{graphane1}).
 The boat conformer has two types of C-C bonds, depending on whether the 
  carbons atoms are bonded to hydrogen atoms which are on the same side
   of the carbon plane or not.
Our calculations indicate that the chair-like conformer is favored by about
$0.051$ eV/atom, in very good agreement with the value of Sofo et al.\cite{graphane1}, who
  found $0.055$ eV/atom. Also, for the chair-like conformer the bond lengths 
  are $1.53$ \AA~(for the C-C bonds) and $1.11$ \AA~(for the C-H bonds), while
   for the boat-like conformer, the C-C bonds are $1.57$ \AA~and  $1.54$ \AA~long,
    and the C-H bonds are $1.10$ \AA~long. These values are essentialy identical
     to the one of Ref. \onlinecite{graphane1}.
 In Fig.\ref{fig:partial} we show the partial density of states (pDOS) for the stable chair conformer, and it is clear that
 the  occupied H states are located over an energy interval starting from the valence band maximum and extending some 10 eV below. 
 As regards to the unoccupied states, the H and C states are also extended and cover a wide energy range.

   \vspace{1cm}

   \vspace{1cm}

\begin{figure}[ht]
 \begin{center}
    \includegraphics[width=0.4\textwidth, angle=0.]{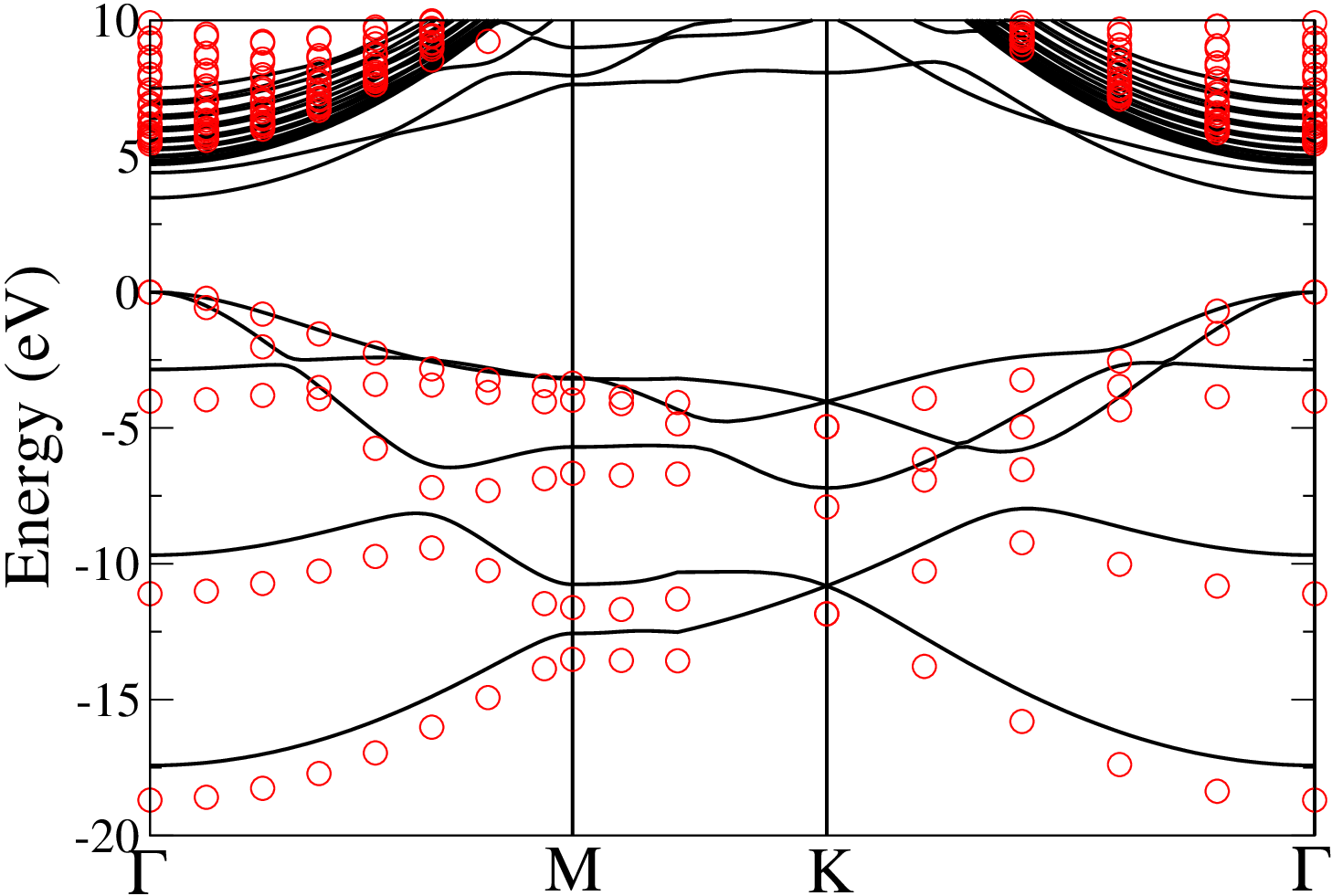}
    \caption{(Color online)
    The GGA bandstructure (full lines) and the GW bandstructure (red dots) 
    of graphane in the chair conformation.
     The top of the valence bands is chosen as the zero energy.
 }
    \label{fig:chair}
  \end{center}
\end{figure}

\begin{figure}[ht]
 \begin{center}
    \includegraphics[width=0.4\textwidth, angle=0.]{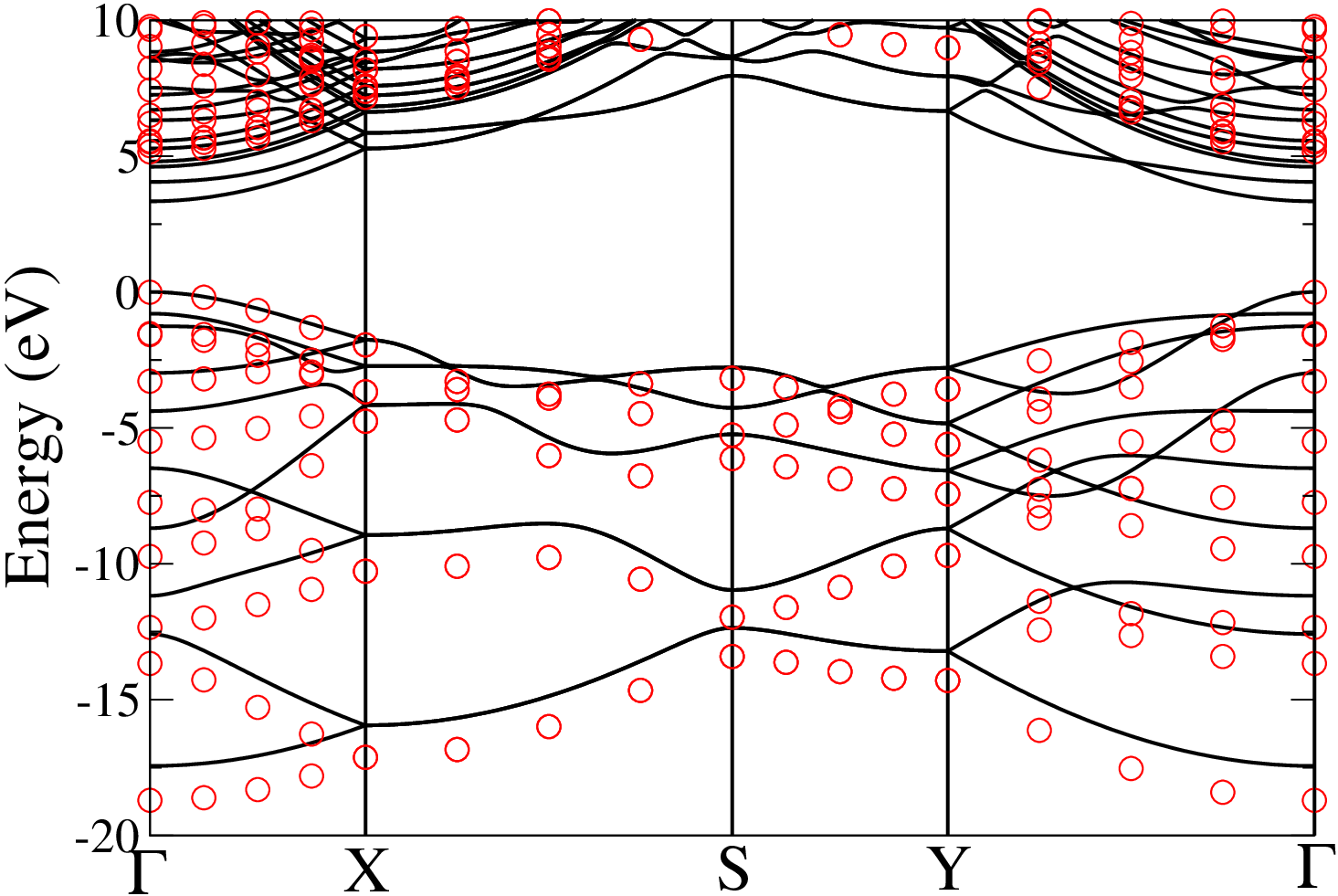}
    \caption{(Color online)
    The GGA bandstructure (full lines) and the GW bandstructure (red dots) 
    of graphane in the boat conformation.
     The top of the valence bands is chosen as the zero energy.
 }
    \label{fig:boat}
  \end{center}
\end{figure}

  Being confident about the reliability of our ground state calculations, we now
   turn to the calculation of excited state properties.
   The GGA (full lines) and GW (dots) bandstructures for both conformers are
   presented in Fig \ref{fig:chair} and \ref{fig:boat}.
     Firstly, our calculated GGA bandstructure for the chair conformation agrees 
      well with the one of Sofo et al (notice that the bandstructure of the boat conformation was not
       reported in this paper).
        The minimum band gap is direct at the high symmetry point $\Gamma$ and has a value of $3.5$ eV.
	 However, this value is dramatically changed when using the more reliable GW approximation,
	  in this case the value is $5.4$ eV. The transitions at the high symmetry points M and K
	   are also significantly increased, from $10.8$ eV and $12.2$ eV with the GGA approximation,
	   to $13.7$ eV and $15.9$ eV with the GW approximation (see Table \ref{tab:tr}).
      For the boat conformation (Fig. \ref{fig:boat}), the band gap is also found to be direct
       at the $\Gamma$ point, with a value given by the GW approximation of $4.9$ eV, whereas
        the GGA gives only $3.3$ eV. The values of the transitions at the X, S, and Y high-symmetry
	 points are also notably corrected, see Table \ref{tab:tr}.
Our calculations show that graphane, in both conformers, can be qualified as a large band gap insulator,
and the electronic properties are expected to be quite different from the ones of
graphene. This is qualitatively in agreement with observations.

\begin{table}
\caption{
\label{tab:tr}
  Values in eV of the transition energies of graphane at some high-symmetry points
   of the Brillouin zone for both conformers (chair or boat). The minimum band gap occurs at
    the $\Gamma$ point.
    The last two lines refer to calculations performed with a $2 \times 2$ supercell, in
     which either an hydroxyl group (OH) or a H vacancy was introduced.
}
\begin{center}
\begin{tabular}{|c|c|c|c|}
\hline
\multicolumn{1}{|c|}{Conformation } &
\multicolumn{1}{|c|}{ Transition } &
\multicolumn{1}{|c|}{ GGA value (eV)} &
\multicolumn{1}{|c|}{ GW value (eV)}
\\\hline\hline
Chair  &  $\Gamma_v \rightarrow  \Gamma_c$  &  3.5 & 5.4 \\\hline
  &  M$_v$ $\rightarrow$ M$_c$ &  10.8 & 13.7 \\\hline
  &  K$_v$ $\rightarrow$ K$_c$  & 12.2 & 15.9 \\\hline
Boat  &  $\Gamma_v \rightarrow  \Gamma_c$  & 3.3 & 5.1 \\\hline
  &  X$_v$ $\rightarrow$ X$_c$  &  7.0 & 9.0 \\\hline
  &  S$_v$ $\rightarrow$ S$_c$  &  10.7 & 13.9 \\\hline
  &  Y$_v$ $\rightarrow$ Y$_c$  &  9.4 & 12.6 \\\hline
Chair+ OH  &  $\Gamma_v \rightarrow  \Gamma_c$  &  3.3 & 5.0 \\\hline
Chair+ H vacancy  &  $\Gamma_v \rightarrow  \Gamma_c$  &  3.7 & 5.4 \\\hline
\end{tabular}
\end{center}
\end{table}

Since graphane is a wide band-gap material it is relevant to ask if defect states can be introduced easily, and manipulated for electronics applications.
 We have here considered two examples of defects in graphane, in the chair conformation: replacement of a H atom by an hydroxyl group (OH), and H vacancies.
  We have calculated the electronic structure of such systems using a $3 \times 3 $ supercell, taking fully into account the relaxation of the
   structure due to the hydroxyl group or the H vacancy. For both cases, a $2 \times 2 \times 1 $ kpoints mesh was used to obtain reliable geometry, and
    a final calculation with $8 \times 8 \times 1 $ kpoints mesh was conducted to obtain well converged density of states.
 
\begin{figure}[ht]
 \begin{center}
    \includegraphics[width=0.4\textwidth, angle=0.]{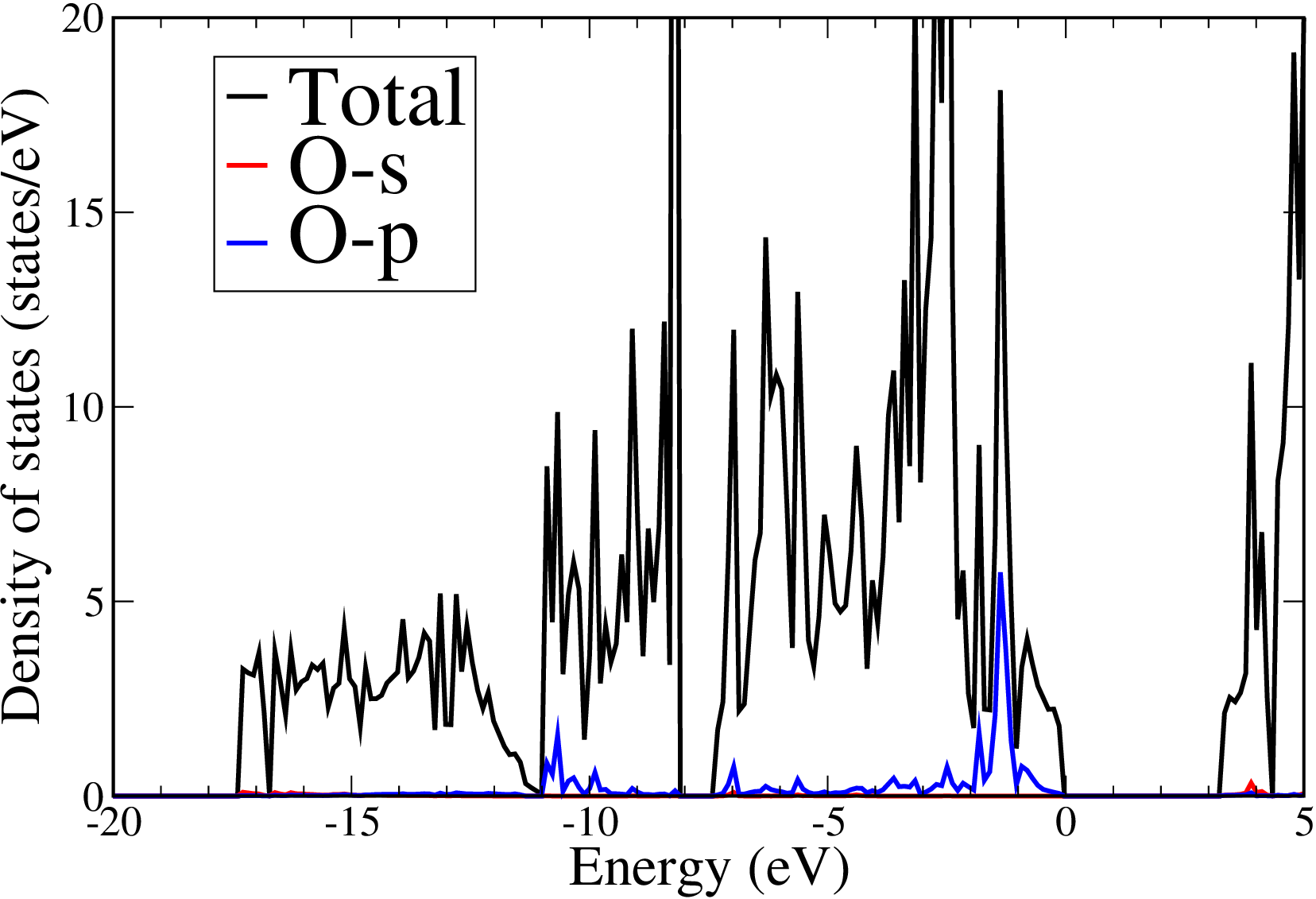}
    \caption{
    \label{fig:defectOH}
    (Color online)
    The total (in black) and partial (in red for O-s states and in blue for O-p states) density of states of
     graphane+OH.
     The top of the valence bands is chosen as the zero energy.
 }
  \end{center}
\end{figure}

    In Fig. \ref{fig:defectOH}, we present our density of states for graphane+OH. The partial density of states of
     oxygen is presented for 's' and 'p'states. It appears that the electronic structure
      of graphane is not significantly modified by the replacement of a H atom by the hydroxyl group. While the O-s 
       states are very widely distributed in energy, the O-p states appear mainly as a peak close to the maximum of
        the valence band, between $-1$ and $-2$ eV.
   \vspace{1cm}
Among the extra electrons brought by the oxygen atom, one electron is involved in a bond with the
  neighbouring C atom, another electron is participating in the bond with H, and the two remaining electrons
   are forming a doublet on the oxygen atom. This is confirmed by the fact that the geometry of the network of
    carbon atoms, driven by sp$^3$ hybridization, is almost unaffected by the addition of the hydroxyl group.
 The band gap (from GGA) is slightly decreased, from $3.5$ eV for
	 pure graphane, to $3.3$ eV here. Here as well, the band gap is underestimated by GGA, so
	  we have applied the GW approximation to get a meaningful value. However, since the numerical
	   cost of the GW approximation is quite high, we had to reduce the size of the supercell to 
	   $2 \times 2 $. In this case, the GW band gap is about 5.0 eV (see Tab.\ref{tab:tr})
	which is in line with the band gap of pure graphane (5.4 eV).

\begin{figure}[ht]
 \begin{center}
    \includegraphics[width=0.4\textwidth, angle=0.]{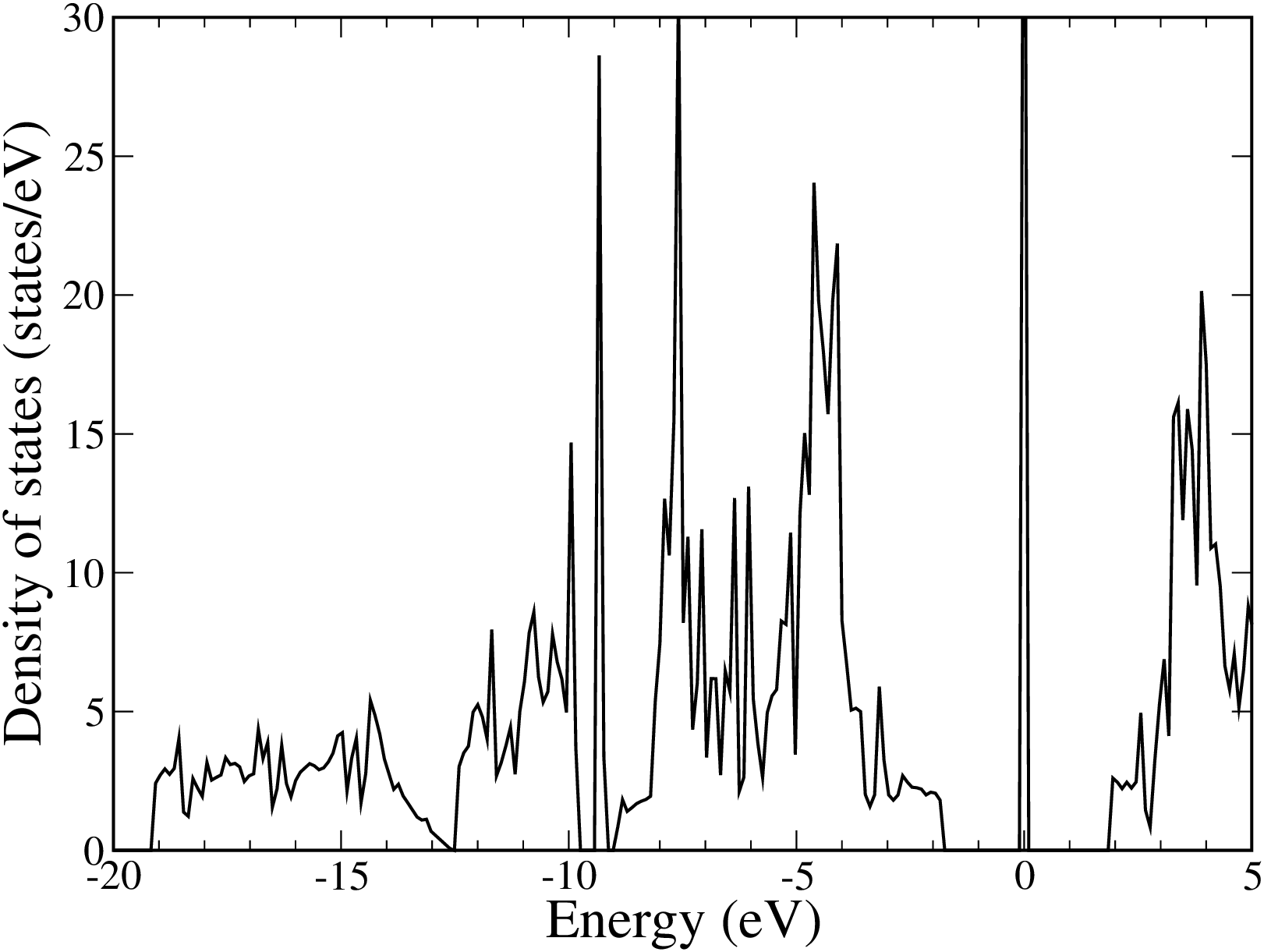}
    \caption{
    \label{fig:defectH}
    The total density of states of graphane+ a H vacancy.
    The Fermi level is put at zero energy.
 }
  \end{center}
\end{figure}

	Next we introduced a hydrogen vacancy in graphane. Then, not every C atom of the graphane layer
       of the graphane layer is saturated with a H atom.
   From inspection of the relaxed geometry, it appears that the bonds between the carbon neighbouring the H vacancy
   and the carbon network have gained a significant part of  sp$^2$ hybridization, like in graphene, with bond lengths here of
    $1.49$ \AA~($1.42$ \AA~in graphene).
  The resulting total density of states is shown in Fig.\ref{fig:defectH}.
   \vspace{1cm}
  An impurity state develops
  close to 2 eV above the valence band maximum, at the Fermi level, changing significantly the overall
   picture of the DOS.
   Also, using the GW approximation (for a $2 \times 2 $ supercell), the difference in energy between the valence band maximum and the conduction
    band mininum is increased from 3.7 eV to 5.4 eV, but the impurity state at the Fermi level is not affected by the GW correction.


To summarize, in this article, results concerning the electronic structure of graphane have been
 presented using the GW approximation. It was found that for both conformers, 
  the minimim band gap is direct and located at the $\Gamma$ point. For the stable chair
   conformation, the most stable one, it has a value of $5.4$ eV, while for
    the boat conformation, it has a value of 4.9 eV. 
  We also found that defects in graphane, in the form of H vacancies, provide an impurity state some 2 eV above the valence band maximum,
  a finding which may be important when trying to use this material for electronics applications.

\begin{acknowledgments}
	S. L. is grateful to G. Kresse for providing the version 5 of VASP
	 and acknowledges financial support from ANR PNANO Grant ANR-
  06-NANO-053-02 and ANR Grant ANR-BLAN07-1-186138.
  M. K. and O.E. acknowledge financial support from Vetenskapsr\aa det (VR), G\"oran Gustafsson Stiftelse, and SNIC/SNAC. 
  M. I. K. acknowledges financial support from FOM (Netherlands).
\end{acknowledgments}



\begin{thebibliography}{49}
\bibitem{first} K. S. Novoselov et al, Science {\bf 306}, 666
(2004).

\bibitem{r1} A. K. Geim and K. S. Novoselov,
Nature Mater. {\bf 6}, 183 (2007).

\bibitem{r2} M. I. Katsnelson, Mater. Today {\bf 10}, 20 (2007).

\bibitem{r3} A. H. Castro Neto, F. Guinea, N. M. R. Peres, K. S. Novoselov,
and A. K. Geim, Rev. Mod. Phys. {\bf 81}, 109 (2009).

\bibitem{klein} M. I. Katsnelson, K. S. Novoselov, and A. K. Geim,
Nature Phys. {\bf 2}, 620 (2006).

\bibitem{ponomarenko} L. A. Ponomarenko et al, Science {\bf 320},
356 (2008).

\bibitem{elias}
D. C. Elias et al. Science {\bf 323}, 610 (2009).

\bibitem{graphane1} J. O. Sofo, A. S. Chaudhari, and G. D. Barber,
Phys. Rev. B {\bf 75}, 153401 (2007).

\bibitem{graphane2} D. W. Boukhvalov, M. I. Katsnelson, and A. I.
Lichtenstein, Phys. Rev. B {\bf 77}, 035427 (2008).

\bibitem{danil} D. W. Boukhvalov and M. I. Katsnelson,
arXiv:0809.5257 (to appear in J. Phys.: Condens. Matter).

\bibitem{godby} R. W. Godby, M. Schluter, and L. J. Sham,
	Phys. Rev. Lett {\bf 56}, 2415 (1986).

\bibitem{wessely}
O.Wessely, M.I.Katsnelson, A.Nilsson, A.Nikitin, H.Ogasawara,
B.Sanyal and O.Eriksson, Phys. Rev. B {\bf 76}, 161402 (2007).

 \bibitem{Hohenberg}
  P. Hohenberg and W. Kohn, Phys. Rev. {\bf 136} (1964);
  W. Kohn and L.J Sham, Phys. Rev. {\bf 140}, A1113 (1965).

\bibitem{Hedin1}
 L. Hedin, Phys. Rev. {\bf 139}, A796 (1965).

\bibitem{Hedin2}
 L. Hedin and S. Lundquist, in {\sl Solid State Physics}, edited by
 H. Ehrenreich, F. Seitz, and D. Turnbull (Academic, New York, 1969),
 Vol. 23, p. 1.





\bibitem{review_gw1}
F. Aryasetiawan and O. Gunnarsson, Rep. Prog.
Phys. {\bf 61}, 237-312 (1998).

\bibitem{review_gw2}
W. G. Aulbur, L. J\"onsson, and J. W. Wilkins, '{\it
Quasiparticle calculations in solids}',
in Solid State Physics; edited by H. Ehrenreich and F. Spaegen, vol {\bf 54}.

\bibitem{review_gw3}
G. Onida, L. Reining, and A. Rubio,
Rev. Mod. Phys. {\bf 74}, 601 (2002).



\bibitem{vasp}
G. Kresse and D. Joubert, Phys. Rev. B. {\bf 59}, 1758 (1999).

\bibitem{Bloechl}
 P.E Bl\"ochl, Phys. Rev. B {\bf 50}, 17953 (1994).

\bibitem{gga}
 J. P. Perdew, K. Burke, and M. Ernzerhof, Phys. Rev. Lett. {\bf 77}, 3865 (1996).

\bibitem{Monkhorst}
 H. J. Monkhorst and J.D. Pack, Phys. Rev. B {\bf 13}, 5188 (1976).

\bibitem{gwkresse}
{M. Shishkin and G. Kresse}
{Phys. Rev. B} {\bf 74}, 035101 (2006).

\bibitem{brice}
{B. Arnaud and M. Alouani}
{Phys. Rev. B} {\bf 62}, 4464 (2000).

\bibitem{seb}
{S. Leb\`egue, B. Arnaud, M. Alouani, and P. E. Bl\"ochl}
{Phys. Rev. B} {\bf 67}, 155208 (2003).

\bibitem{louie1}
{Y-W Son M. L. Cohen and S. G. Louie}
{Phys. Rev. Lett.} {\bf 97}, 216803 (2006)

\bibitem{louie2}
{L Yang, C-H Park, Y-W Son, M. L. Cohen, and S. G. Louie}
{Phys. Rev. Lett.} {\bf 99}, 186801 (2007)

\bibitem{gwgraphene}
{P. E. Trevisanutto,C. Giorgetti,  L. Reining, M. Ladisa and V. Olevano}
{Phys. Rev. Lett.} {\bf 101}, 226405 (2008)

\bibitem{gwtubes}
{T. Miyake and S. Saita}
{Phys. Rev. B} {\bf 68}, 155424 (2003)







\end{thebibliography}
\end{document}